\documentclass[acmsmall,screen]{acmart}

\AtBeginDocument{\providecommand\BibTeX{{\normalfont B\kern-0.5em{\scshape i\kern-0.25em b}\kern-0.8em\TeX}}}

\begin{document}

\setcopyright{cc}
\setcctype{by}
\acmJournal{PACMHCI}
\acmYear{2026} \acmVolume{10} \acmNumber{2} \acmArticle{CSCW040}
\acmMonth{4} \acmPrice{} \acmDOI{10.1145/3788076}

\title{To Tango or to Disentangle? Making Ethnography Public in the Digital Age}

\author{Daniel Mwesigwa}
\email{dm663@cornell.edu}
\orcid{0000-0002-2735-4875}
\affiliation{\institution{Cornell University}
  \streetaddress{Gates Hall, 107 Hoy Rd}
  \city{Ithaca}
  \state{NY}
  \country{United States}}

\author{Cyan DeVeaux}
\email{cyanjd@stanford.edu}
\orcid{0000-0003-2655-9841}
\affiliation{\institution{Stanford University}
  \streetaddress{Building 120, Room 110 450 Jane Stanford Way}
  \city{Stanford}
  \state{California}
  \country{United States}}

\author{Palashi Vaghela}
\email{palashi_vaghela@sfu.ca}
\orcid{0000-0002-1537-1635}
\affiliation{\institution{Simon Fraser University}
  \streetaddress{250 -13450 102 Avenue}
  \city{Surrey}
  \state{British Columbia}
  \country{Canada}}

\begin{abstract}

Ethnography attends to relations among people, practices, and the technologies that mediate them. Central to this method is the duality of roles ethnographers navigate as researchers and participants and as outsiders and insiders. However, the rise of digital platforms has introduced new opportunities as well as practical and ethical challenges that reshape these dualities across hybrid media environments spanning both online and offline contexts. Drawing on two case studies of VRChat and WhatsApp, we examine how ethnographers employ diverse tactics to study both enduring and emerging socio-cultural issues of race and caste, particularly those that form what are often called publics. We propose \textit{emergent relationality} as a key analytic for understanding the mutual shaping of ethnographers, platforms, and publics. In this work, emergent relationality offers registers for analyzing how positionality and hybrid media environments constitute and condition what can be accessed, articulated, and made public.



%
 \end{abstract}



\maketitle

\section{Introduction}

Ethnography’s commitment to in-depth, real-time, and immersive participant observation has generated real and consequential insights about how social worlds are made, sustained, and transformed \cite{geertz_thick_1973,van_maanen_tales_2011,visweswaran_fictions_1994}. In technology contexts, this has meant sustained attention to design and use as sites where meaning, practice, and social order are negotiated \cite{dourish_implications_2006,irani_postcolonial_2010,kang_towards_2025,khovanskaya_reworking_2017,randall_ethnography_2021,taylor_ethnography_2010,bardzell_towards_2011}. However, ethnography now faces growing practical and methodological concerns due to the shifting configurations of technology, including how these changes affect the formation of social relations and the spaces in which issues of concern emerge \cite{olson_reading_2014,rifat_putting_2022,soden_evaluating_2024,crabtree_h_2024,varanasi_accost_2022}.
Despite these technology-inflected shifts, ethnography's central epistemological and methodological contribution is that of “participant observation,” a method characterized by what anthropologists call “undesigned relationality” \cite{bell_problem_2019}: the ethnographer's identity and relationships are not fixed in advance but emerge through fieldwork. This fluidity is often understood through the duality of roles occupied by ethnographers who are simultaneously observers and participants, outsiders and insiders. While ethnographic practice in computer-supported cooperative work (CSCW) and human-computer interaction (HCI) has historically emphasized the researcher's observational distance \cite{dourish_implications_2006}, a growing number of critical scholars are arguing for a more ‘subjective’ and nuanced approach of blurring the distinctions between the received categories of ‘observer’ (researcher) and ‘participant’ (research subject) \cite{joshi_who_2024,mcdonald_reliability_2019,williams_theres_2010,rode_commentary_2026}. By leveraging this subjectivity and duality as a productive site of inquiry, ethnographers not only reflect on their personal biases and intellectual commitments but also produce accounts from the very relationships of their dual roles as observers and participants and as outsiders and insiders. This then offers a more nuanced and arguably fuller picture of the different interpretive frames from which field actors impute meaning.

While the ethnographic study of the internet and digital platforms has compared and contrasted interactions in digital spaces and physical spaces \cite{dourish_re-space-ing_2006}, it is now perhaps clearer that “space is now permanently both physical {\em and} virtual” \cite{small_ethnography_2022} (emphasis in original text). The often taken-for-granted social, cultural, and political aspects of everyday life simultaneously permeate both online and offline worlds. For instance, the continued rise and adoption of digital platforms such as virtual reality (VR) and social networking have facilitated new hybrid relations, destabilizing the boundaries between the physical and virtual \cite{boellstorff_ethnography_2024}. These digital platforms are also “inherently entangled [...] socio-technical assemblages”  \cite{gerlitz_apps_2019}, which simultaneously serve as bases for social and political action \cite{plantin_infrastructure_2018,star_steps_1994}. As such, these platforms double as digital infrastructures where broad and narrow forms of participation cultivate or foreclose the formation of publics \cite{marres_infrastructural_2025}. Prior work in CSCW and adjacent fields has highlighted the central role played by infrastructures in conditioning publics; particularly the articulation and negotiation of issues of concern \cite{disalvo_design_2009, jackson_hijacking_2015,le_dantec_tale_2010,dantec_infrastructuring_2013,ludwig_publics_2016}. 
Across CSCW and HCI venues, publics have primarily been drawn on to understand deliberative democracy \cite{nelimarkka_review_2019,nelimarkka_what_2024} and also as a resource for participatory design \cite{disalvo_design_2009,dantec_infrastructuring_2013}. 
However, recent scholarship goes beyond platform-centric publics (Cf. \cite{gillespie_relevance_2014,boyd_white_2011,tufekci_engineering_2014,christin_multiple_2020}) to account for the hybrid media environments that span online and offline contexts through which technology mediates everyday life,  social relations and issues of concern \cite{moller_hartley_researching_2023}. What might ethnography offer to the study of publics in the digital age, and how might attending to publics transform ethnographic practice?

The paper's main contribution is \textbf{\textit{emergent relationality}}, a key analytic we have developed to connect the practice of ethnography to the study of publics via platforms and in real-world contexts. Emergent relationality is not a method but is composed of registers of observation that support ethnographers in examining relations and issues of concern across hybrid media environments. Using emergent relationality, we reflect on two cases to read into the material and discursive formation of publics. The first case reads into issues of racial representation in VR, showing how Black experiences in VRChat are cultivated not only for community but also play, despite the risk of racially-charged harms. The second case study reads into the silences written in supposedly ‘casteless’ worlds of computing by examining how WhatsApp, a popular social messaging app, renders caste across the app's affordances of group and direct messages. Leveraging emergent relationality, we argue that the ethnographer's dual role as observer and participant, outsider and insider, is not separable from the sites through which fieldwork proceeds and the publics that emerge in practice. 
Emergent relationality therefore foregrounds how the ethnographer's positionality and hybrid media environments constitute and condition access and what can be articulated within them.

The following sections begin with a review of CSCW, HCI, and social science literature on the performative nature of ethnography, with particular attention to the \textit{multiple} dual roles ethnographers occupy as they navigate and experience the field. We then describe infrastructural participation and its constituent elements, platforms and publics, which are bridged through ethnographic practice. Drawing on this literature, we develop and introduce emergent relationality, the key analytic that informs our reflections on two case studies situated within \textit{closed} social media platforms, VRChat and WhatsApp. Our discussion explores how emergent relationality opens up new ways of understanding platforms and publics as a relational landscape, while also recognizing the affective work undertaken by researchers across varied field settings. Finally, we outline methodological implications for ethnographic practice and design research.

\section{Performativity of (digital) ethnography}

Ethnography is “intrinsically performative” \cite{bell_problem_2019}. As such, ethnography is not merely about collecting ‘field data’ but an active process of \textit{making} research data, enriched through engagement with broader lines of social theory and sustained reflection about field practice \cite{dourish_implications_2006,joshi_who_2024,rapp_reflexive_2018,upadhyay_positionality_2024,lazem_challenges_2022}. As part of the tools of the trade, the ethnographic eye and full range of senses are tuned to the ambivalent, silenced, and even familiar, in part adopting Tsing's “arts of noticing [...] to listen [and see the human and non-human] and tell a rush of stories” \cite[p. ~37]{tsing_mushroom_2015}. At the same time, ethnography shifts the idea of a `site' from a fixed, bounded space by constructing the field-site(s) as a dynamic, heterogeneous space where social phenomena are closely tracked \cite{burrell_field_2009}.  

Although observational distance has been emphasized in practice \cite{dourish_implications_2006}, the dynamic and entangled roles that ethnographers occupy are the foundation of the method.
Anthropologist Kirsten \citet{bell_problem_2019} uses “undesigned relationality” to illustrate that the ethnographer's identity and relationships emerge through fieldwork rather than being predetermined. Procedural ethics frameworks, rooted in what she calls the “Standard Model,” treat this as a problem to be managed. But for the ethnographic traditions this paper draws on, undesigned relationality is not a problem; it is the foundation of the method, which also supports ethical practice. 
The boundaries between observer and participant, outsider and insider, online and offline become not merely blurred but multiply configured. These configurations shape field access and how the field site is constructed and experienced across hybrid media environments. In virtual worlds, for example, research and play commingle \cite{boellstorff_ethnography_2012}: understanding the mechanics of play is prerequisite to meaningful engagement with virtual cultures \cite{gray_race_2014}. In foundational workplace studies, ethnographers were not just `external' observers but also `insiders' embedded in the same distributed milieux as the scientists they studied \cite{suchman_human-machine_2006,star_ethnography_1999}. Yet as Black feminist scholar Patricia Hill Collins \cite{collins_reflections_1999} has valuably suggested, the hybrid “outsider-within” roles which emerge do not eliminate the power structures that continuously shape field work. Social locations of race, gender, and class persist even as other boundaries become fluid.

In CSCW and social informatics, ethnography is alive to these dynamics, and the increasing demands for methodological rigor and ethical practice. Renewed thinking about the politics and ethics of feminist ethnographic research has generated critical insights about centering participant voices in the mutual production of knowledge and questioning of power asymmetries between researchers and participants \cite{liang_embracing_2021,bardzell_towards_2011,ajmani_moving_2025}, as well as in subverting dominant viewpoints \cite{ogbonnaya-ogburu_critical_2020, vaghela_interrupting_2022, vaghela_birds_2021, vaghela_hidden_2023}. In doing so, ethnographers have told ``fewer lies'' \cite{de_seta_three_2020,fine_ten_1993} by building on the perspectival nature of practice (i.e., the ethnographer's point of view) and reflexivity, the confessional practice used to account for personal biases, assumptions, and other commitments \cite{joshi_who_2024,bardzell_towards_2011,rapp_reflexive_2018,williams_theres_2010}. While this renewed thinking has fostered stronger practical and ethical commitments to research transparency and rigor, ethnography is beset by questions of access, positionality, and scale in the digital age. For instance, the dynamics of “informed consent” have shifted in an era of pervasive data research techniques, where it has become much easier to collect and analyze expansive and disconnected datasets \cite{shilton_excavating_2021}. Although IRB exempts such ``public data'' \cite{shilton_excavating_2021,zimmer_but_2010}, informed consent still constitutes ongoing ethnographer negotiations and performances meant to manage field access, while minimizing possibilities of harm. As \citet{visweswaran_fictions_1994} reminds us in \textit{Fictions of Feminist Ethnography}, consent is never unambiguous, and moments of refusal signal not a breakdown of method but the ethical limits of inquiry.

Altogether, even though ethnographies are \textit{collaboratively} accomplished, the ethnographer is the principal author of their output. Yet the performative nature of ethnography makes it both an evolving process and the product of relationships within and with the field sites and new modes of participation. How ethnographers perform this relational work as publics emerge and are negotiated in hybrid media environments remains undertheorized.

\section{Infrastructural participation} \label{acc}

It is now banal to point out that algorithmic systems (or information infrastructures or digital platforms or whatever they will be called by the time you read this) mediate social, economic, political, and cultural spheres of everyday practice, in which they are also (re)constituted. As several CSCW and STS scholars have shown, interfacing or encountering such computational systems is now inescapable to a large extent, knowingly or unknowingly \cite{singh_seeing_2021,dourish_algorithms_2016,baumer_algorithmic_2024}. \citet{marres_infrastructural_2025} have framed this condition as \textit{infrastructural participation}, loosely defined as the structured and mundane activities that people perform and the webs of relations they weave towards some activity or goal, often involving digital platforms. Across four empirical sites, the authors summarize two forms of infrastructural participation, technology-centric and society-centric. Drawing on mobility and ed-tech platforms they show how technology-centric participation tends to valorize user engagement, data accumulation, and long-term profit while society-centric participation cultivates distinctive and heterogeneous interests of society, particularly in domains of citizen science and data privacy. The analytical and methodological focus on both infrastructure and publics illustrates how these two elements are deeply entangled, and why they are a principal {\em space} and {\em condition} for participation \cite{baringhorst_how_2019}. Below we examine the two elements of infrastructural participation (platforms and publics) before showing how ethnography bridges them.

\subsection{Platforms as infrastructures}
Scholarship in CSCW and media studies has developed and extended the notion that digital platforms {\em are} infrastructures \cite{plantin_infrastructure_2018,dourish_hci_2010,gerlitz_apps_2019}, highlighting how the term `platform' is only a discursive shift meant to serve varying technical, political, legal, and cultural interests \cite{gillespie_politics_2010}. Like other infrastructure, platforms serve as bases for social and political action \cite{plantin_infrastructure_2018}. Platforms also position their messaging as benign intermediaries with promises to various groups of \textit{users} \cite{gillespie_politics_2010}: they ``level the ground'' for small content creators; provide targeted social listening and geo-demographic tools to advertisers; while also appealing to protect the strategic interests of incumbent media industries. Although mainstream platforms are often owned and managed by private corporations, there have been growing partnerships between private platforms and state actors, which has expanded the infrastructural reach and structuring power of digital platforms. Accordingly, platforms are anything but neutral \cite{gillespie_platforms_2018,gillespie_politics_2010}. For instance, Google's educational offerings are powering public education in Europe \cite{marres_infrastructural_2025}, while Facebook is often \textit{the} internet in many parts of the world \cite{nothias_access_2020}. 

Platforms also come with distinct (if evolving) \textit{points of view}. James Scott's work \textit{Seeing Like a State} \cite{scott_seeing_1998} elaborated the efforts by states to describe and represent their citizens and borders in legible registers of control. A powerful metaphor of vision, ``seeing like'' allows us to understand different viewpoints within and of digital platforms, something CSCW and HCI scholars have already been doing \cite{singh_seeing_2021,dourish_seeing_2007,sambasivan_seeing_2021}. In \textit{seeing like an interface}, \citet{dourish_seeing_2007} argues that information systems act as new sites of spatial experience, connecting people across distributed patterns of activity. In VR, for example, platforms dramatically expand how users engage and connect \cite{nardi_my_2010,gray_race_2014}. For social messaging platforms like WhatsApp, communication possibilities have transcended the personal into arenas such as commerce \cite{joshi_reselling_2025,de_business_2025} and politics \cite{varanasi_accost_2022,shahid_one_2025}. However, while platforms are embedded and constituted in socially meaningful encounters between people and places, there are always constraints and tensions emerging from what a platform sees and the responses its modes of seeing elicit from a range of stakeholders \cite{gilbert_towards_2023,plantin_infrastructure_2018,karizat_algorithmic_2021,nunes_algorithmic_2024,de_business_2025}.

\subsection{Publics and their interpretations}
Popular interpretations of publics in CSCW, HCI, and design studies have tended to dwell on pragmatist (e.g., \cite{dewey_public_2016, marres_material_2015,latour_realpolitik_2005}) and symbolic interactionist (e.g., \cite{fine_tiny_2012}) interpretations, which frame publics as emergent through interaction, inquiry, and shared issues rather than as pre-existing social groups or formal constituencies. Most work follows American pragmatist philosopher John Dewey \cite{dewey_public_2016}, who has articulated how collectives are “called into being” not only by problems and events of interest but also by the materiality of infrastructural (and indeed platform) connections linked by “vast currents” of issues \cite{collier_limn_2016,marres_material_2015,dantec_infrastructuring_2013}. Interactionist sociologist Alan Fine \cite{fine_tiny_2012} has developed ``tiny publics'' to show how the actions of “small groups” can generate civic engagement and forge shared identities, even without particular attention to \textit{specific} issues (\textit{pace} Dewey). These interpretations have inspired an important body of interventionist critical design research and participatory work grounded in the re-imagination of “new social, economic, and political arrangements” \cite{disalvo_making_2014}. For example, LeDantec and DiSalvo's publics work encapsulates the development participatory sensing platforms, neighborhood issue-tracking tools, and community data infrastructures \cite{dantec_infrastructuring_2013,disalvo_making_2014}. For \citet{steup_growing_2018}, disparate groups of small holder farmers in the US, organized across distributed needs and concerns of local food movements have forged multiple tiny publics (despite a coherent articulation of a singular issue) to support them in challenging the excesses of global food supply chains. Beyond design intervention-driven publics, scholars in CSCW and allied fields including media studies and STS have suggested that artifacts have agency, where they actively participate and do things in the world. \citet{jenkins_object-oriented_2016} call this ``object-oriented publics.'' Examples of such publics might include: \citet{tufekci_engineering_2014}'s ``engineered publics'' (how platforms direct what is seen); \citet{christin_multiple_2020}'s ``algorithmic publics'' (how algorithms structure audiences); \citet{boyd_white_2011}'s ``networked publics'' (how properties of networked technologies shape social networks); \citet{gillespie_relevance_2014}'s ``calculated publics'' (how classifications of user behavior become socially meaningful). 

A parallel but related line of work is concerned with the broad \textit{spheres} in which publics are animated, and the tensions that arise as publics take shape. \citet{baym_sonic_2019} have shown how previously utopian technologies such as blockchain have been reframed as a ``convening technology,'' thus structuring a ``sonic public'' that brings together different groups of actors in the music industry. More recent technologies such as generative AI have prompted the rise of similar publics, albeit under names such as ``AI publics'' \cite{sieber_who_2024} and ``experimental publics'' \cite{metcalf_experimental_2025}. At the heart of these spheres is controversy and tension amongst groups brought together under different convening technologies. But as several critical scholars have shown -- in earlier work critiquing Jürgen Habermas’s \cite{habermas_structural_1990} formulation of the public sphere as site of rational debate -- these publics elide the majority: who can convene and influence debate, at whose expense? \cite{baym_sonic_2019,amrute_thinking_2025}. The thinking on ``counter-publics'' emerged in part to articulate the issues (through the affective and creative work) of subaltern communities such as women, racialized people, and sexual minorities \cite{fraser_rethinking_1990,warner_publics_2002,jackson_hijacking_2015,jackson_recentering_2023,le-khac_blm_2022}. 

While publics are emergent and multiple, they are empirically challenging to study and fuzzy to fully conceptualize. However, the relationality of publics --  “open-ended networks of actors (i.e., without a closed or fixed membership) linked together through flows of communication, shared stories, and civic or other collective concerns” \cite[p. ~69] {starr_relational_2021} -- offers \textit{the} interpretive flexibility to apprehend and interpret publics in increasingly hybrid media environments \cite{moller_hartley_researching_2023}. At the same time, because of this relational grounding, publics can then be understood differently depending on whether they are viewed from the perspective of ethnographers or platforms, or both.

\subsection{Ethnographic bridge between platforms and publics}
Platforms and publics are mutually constituted. Indeed, digital platforms have not only become sites where publics emerge, but also where users and other stakeholders wrestle with digital platforms over rights and responsibilities for what should constitute technical, political, and aesthetic order \cite{dourish_allure_2021,kelty_geeks_2005}. Platforms and publics, thus, offer compelling sites for ethnographic inquiry and reflection. Because ethnographers are trained to be sensitive to multiple interpretations and attend to silences, norms, and values \cite{star_ethnography_1999}, their navigation of differential access as researchers \textit{and} participants offers a unique perspective into how they see how publics are infrastructured in practice \cite{singh_seeing_2021}. Learning to see like an infrastructure might, for instance, entail auto-ethnographic observations of a platform as well as reading the inscriptions written in platform affordances \cite{bucher_technicity_2012}. Perhaps by what \citet{star_ethnography_1999} has termed ``following,'' we can turn the notion of \textit{making} field data on its head -- by following the material artifacts constructed by people (e.g., posts), trace or log of user activity (browsing patterns), and the changing rules and policies that shape content (platform governance), among others. 

CSCW researchers have ethnographically studied and followed actors, actions, and artifacts across field sites using two broad and interrelated approaches: \textbf{computational ethnography} (machine-readable) and  \textbf{ethnography on/of computational platforms} (human-readable). First, computational ethnography is a mixed-method approach that takes advantage of classic ethnographic sensibilities and computational tools and tactics of social science. A popular example of computational ethnography is “trace ethnography” \cite{geiger_trace_2011}, involving the study of the “everyday affairs of a group and the active investigation of actors, software and data that otherwise exist in the background.” Through the use of computational tools such as APIs (application programming interfaces), trace ethnography enables researchers to “interview” databases about what platforms “observed” \cite{geiger_trace_2011}. Second, ethnography on digital platforms is a predominantly qualitative research practice that closely follows the principles and techniques used in classic ethnography, including translating concepts such as “participant observation” to the interpretive study of sociocultural action in digital arenas. Techniques here might include ad-hoc ethnographic observations and entail manual data collection “(i.e., screengrabs, copy/pasting)” \cite{proferes_studying_2021}. Using these approaches on platforms such as Reddit and TikTok, ethnographers have studied various topics, including mental health \cite{eagle_exploring_2022,feuston_everyday_2019,yildiz_examining_2019}, harmful information \cite{mcclure_haughey_bridging_2022}, and collaboration practices \cite{de_souza_santos_grounded_2022,yuan_tabletop_2021}.

Taken together, the use of ethnography in the study of platforms and publics is informed by the ethnographer’s “arts of noticing,” as well as by attention to the multifarious ways hybrid media environments structure and condition publics. These practices draw on both computational and qualitative approaches, enabling ethnographers to navigate increasingly complex, data-saturated, and algorithmically mediated worlds in which new questions and practical concerns continually emerge in everyday life. While platforms and publics have each been extensively studied in CSCW and allied fields, they are most often examined as analytically distinct objects. Ethnographic practice, by contrast, routinely spans and entangles the two, yet lacks a shared analytic vocabulary for naming and theorizing this work.

\section{Emergent relationality} \label{emerg-rel}

\begin{figure}[htbp]
    \centering
    \includegraphics[width=0.8\textwidth]{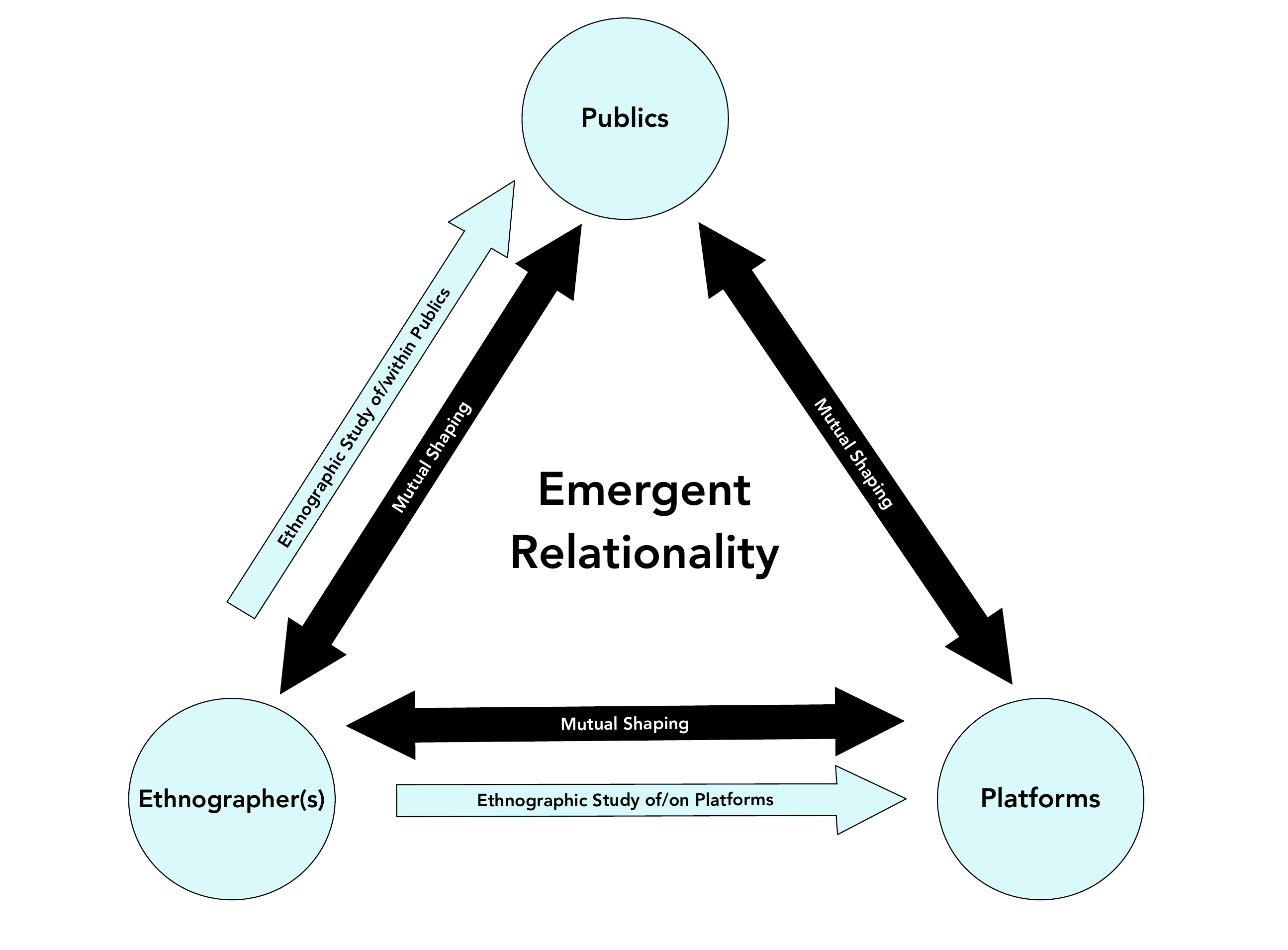}
    \caption{The entanglement of ethnographers, platforms, and publics (in hybrid media environments)}
    \Description{The entanglement of ethnographers, platforms, and publics (in hybrid media environments}
    \label{fig:er_fig}
\end{figure}

The concept of emergent relationality developed here brings out and extends insights from the scholarship above. In relation to the performativity of (digital) ethnography, enduring questions around ethics and practice of ethnography bring out specific tensions and possibilities, which are multiplied by the duality of roles as observer/participant and outsider/insider in hybrid media environments. This not only foregrounds the ethnographer's ‘point of view’ but also enriches ethnographic stories through enhanced reflection and methodological rigor. In relation to infrastructural participation, the range of actors (and their roles) and spatial references (across offline/online boundaries) dramatically expand how the `field' is constructed and experienced. While platform-specific “points of view” condition publics by facilitating collective action and belonging across distances, identities, and cultures \cite{schmidt_constructing_2013,seering_applications_2018}, the issues that animate these publics routinely spill across hybrid media environments. The emergence of publics is therefore entangled in relationships among actors and artifacts, taking shape through networks formed by individuals, including research participants and ethnographers, and by platforms, including their affordances and governance rules. Emergent relationality attends to \textit{access}: its conditions, costs, and limits within hybrid media environments. Who gains entry, on what terms? What can be seen and said once inside? Taken together, emergent relationality attunes ethnographic practice to the mutual shaping of ethnographers, platforms, and publics. It foregrounds the reflexive and tactical work this entanglement demands across hybrid media environments. 
In the following paragraphs, we show how this analytic works (see Figure~\ref{fig:er_fig}).

Although the ethnographer is immersed in the platforms-publics nexus (constituted within infrastructural participation), they still hold a privileged position as the principal author of the ethnographic record. As such, ethnographic attention to both publics and platforms is {\em uni-directional}, originating from the ethnographer who simultaneously observes publics and platforms (using hybrids of machine- or human-readable methods). The ethnographer’s sense- and meaning-making abilities are as much matters of skill as they are of improvisation. For example, an ethnographer's knowledge plays an important role in understanding how a platform's design and affordances, including how platforms could shape publics, and the other way around. At the same time, the improvisation in-situ helps them read against abundances or ``silences''  in the field \cite{soden_time_2021,rettberg_situated_2020}. 

Yet there are also {\em bi-directional} relationships between ethnographers, platforms, and publics, suggesting that each of these entities influence each other through differing interactional capacities, whether the ethnographer recognizes such capacities or not. To be sure, the ethnographer is still the principal author even within the perspective. If the uni-directional relations (above) highlight the ethnographer's skill and improvisational capabilities, the bi-directional relations in this case (illustrated in black arrows) bring to the fore issues of scale and opacity, which complicate how data can be ``followed'' and interpreted. Attending to issues of scale in these contexts is challenging because of the ``black box'' nature of how platforms (and algorithms) function, and their effects on the interactions of users (ethnographers and participants) \cite{christin_algorithmic_2020}. However, instead of a unilateral approach to attempting to open the proverbial ‘black box’ by revealing its underlying mechanisms of operation (an near impossible task), more efforts are redirected towards understanding the \textit{work} that black-boxing performs \cite{marda_importance_2021}, and naming that work appropriately. For example, the relationship between publics and platforms can be understood through lenses of ``object-oriented publics'' whereas an ethnographer's respective relationships with publics and platforms can be understood through the actions of the ethnographer and their visible or invisible effects. In this way, scale and opacity are not constraints \textit{per se} but resources to be engaged in apprehending the fragmented and incomplete nature of \textit{reality} (and data) in hybrid media environments. Altogether, the uni- and bi-directional relations support the knowledge, experience, and improvisational capabilities of an ethnographer in navigating differential access to constitutive parts of fieldwork -- publics and platforms -- while also remaining alert to the fundamental limits of data collected (and made) in these hybrid media environments.

\section{Case studies}

Using \textit{human-readable} ethnographic methods, we develop and reflect on two case studies to illustrate emergent relationality in practice. With a nod to \citet{joshi_who_2024}, we also build on John Van Maanen’s foundational \cite{van_maanen_tales_2011} \textit{Tales of the Field: On Writing Ethnography} to develop impressionist accounts and reflections of our field experiences. According to Van Maanen, impressionist tales, which are a subgenre of ethnographic writing, render special moments to convey highly personalized perspectives. He adds: “Impressionist tales present the doing of fieldwork rather than simply the doer or the done. They reconstruct in dramatic form those periods the author regards as especially notable and hence reportable” \cite[p. ~102]{van_maanen_tales_2011}. By reflecting on our past field experiences (and published work: e.g., \cite{deveaux_black_2025,vaghela_interrupting_2022}), we analyze and reflect on our roles as ethnographers and our cultural projects of inquiry. In the first case study, using words and imagery, we reflect on social play and belonging for Blackness in VR spaces; whereas in the second case study, we reflect on the accounts of castelessness of Indian women in computing across social messaging groups. 

Digital platforms are central to our case studies. We use VRChat and WhatsApp as principal field sites through which we engage the concept of emergent relationality. It is important to highlight that ethnographic research across more public platforms such as Reddit and Twitter (as we have already shown in section \ref{acc}) has produced interesting insights on a broad range of topics. However, less public platforms such as VRChat and WhatsApp have only recently begun to attract the attention of CSCW and HCI researchers: most of this attention stems from the limits of these platforms amidst their growing popularity. \textbf{VRChat} is a growing social VR platform that facilitates immersive digital interactions. Through novel avatar-mediated interactions, users' physical bodies become the direct interface between the physical and digital \cite{freeman_body_2021}. \textbf{WhatsApp} is a social messaging platform that is almost similar to traditional SMS, as both enable one-on-one messaging, broadcasts, and group communication. However, WhatsApp differs because it supports media-rich content and uses end-to-end encryption. WhatsApp’s affordances have been spotlighted for inhibiting content moderation while fostering the spread of harmful content \cite{shahid_one_2025,agarwal_conversational_2024,varanasi_accost_2022}.

Based on ethnographic vignettes and reflections around the platforms identified above, we explore each of these in our IRB-approved case studies. All names used in our stories are pseudonyms.

\subsection*{Of embodied publics in VRChat (Author of first story)}
Between January and September 2022, I conducted ethnographic research on VRChat, a social VR platform. At the time, social VR was experiencing growth in its user base due to the pandemic and the increasing accessibility of commercial VR headsets. I chose social VR as a field site because it represented an emerging sociotechnical context where interactions and digital representations differ experientially from those on traditional platforms. Through head-mounted displays and motion tracking, users experience VR environments through the eyes and body of their avatars. This setting offered a unique lens for exploring the nuances of identity practices and culture formation in immersive virtual worlds. 

To become an ethnographer of VRChat, it was critical to also become a user of the platform. Prior work has outlined that it can be difficult to study technocultural activity in virtual worlds without some degree of involvement in them \cite{boellstorff_ethnography_2024}. To fully understand the social VR experience, I needed to engage with its fundamental components: the ‘social’ and the ‘virtual reality.’ However, it quickly became evident that my engagement with each would be shaped by who I am in real life, not just as a researcher but as a person with lived identities that were reflected in my digital persona. 

\begin{figure}[t]
    \centering
    \includegraphics[width=0.48\linewidth,height=4.2cm]{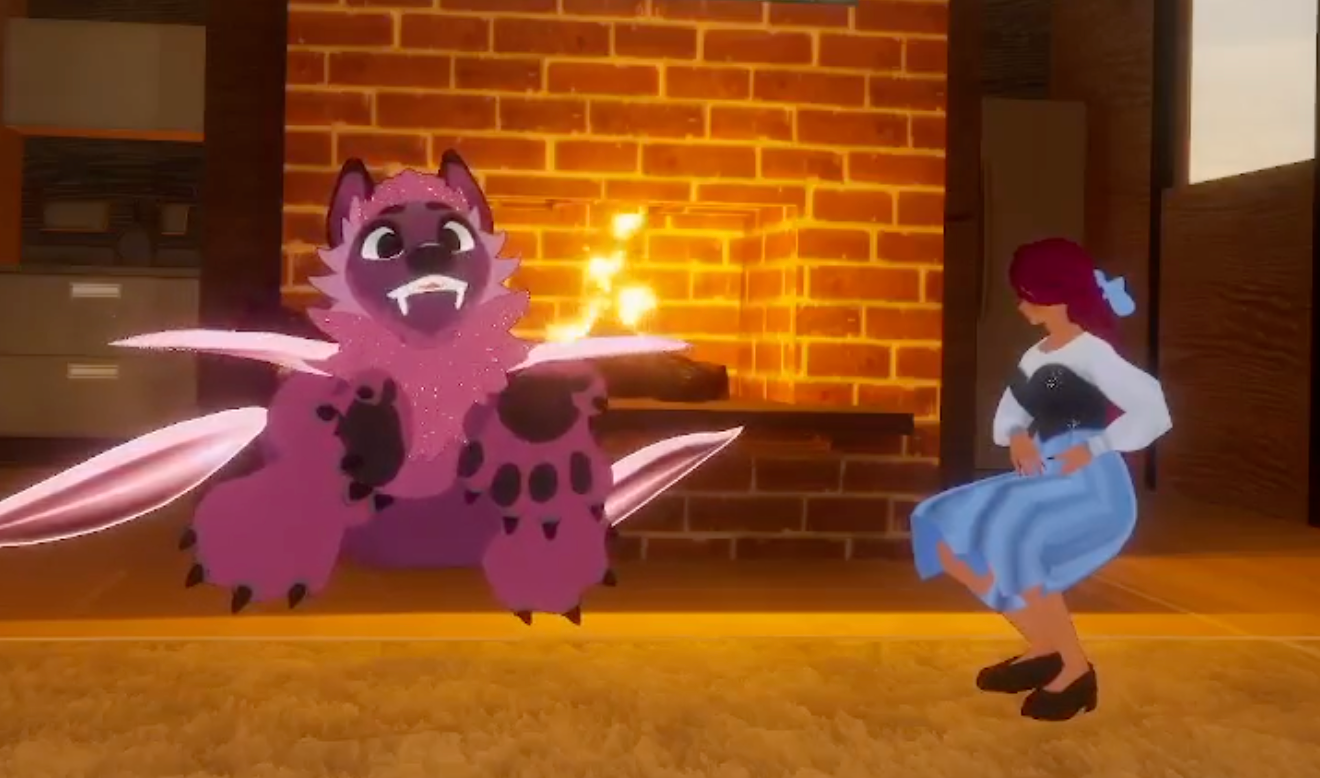}\hspace{4pt}\includegraphics[width=0.48\linewidth,height=4.2cm]{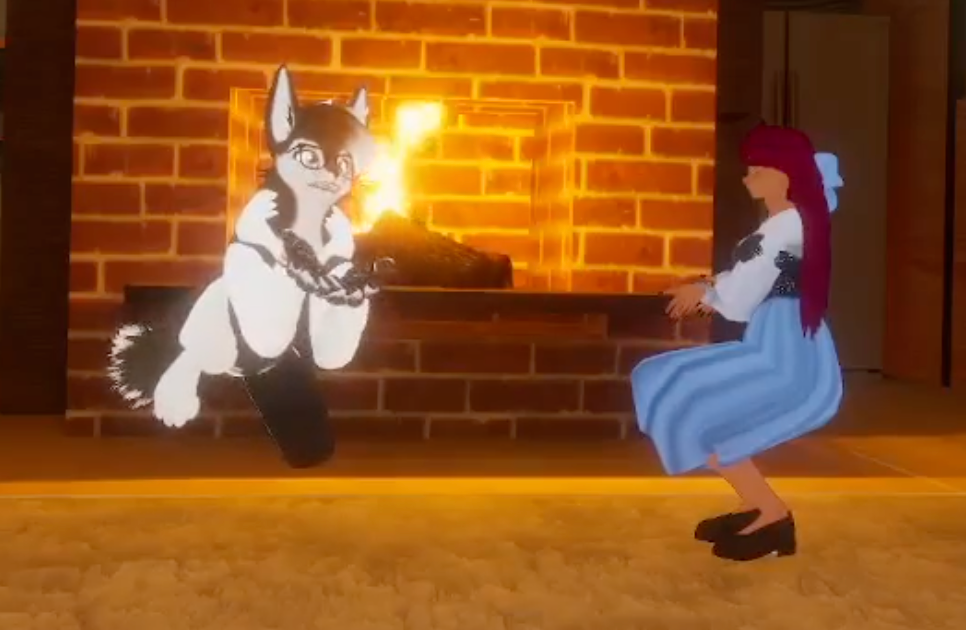}\par\nointerlineskip\vspace{4pt}
    \includegraphics[width=0.48\linewidth,height=4.2cm]{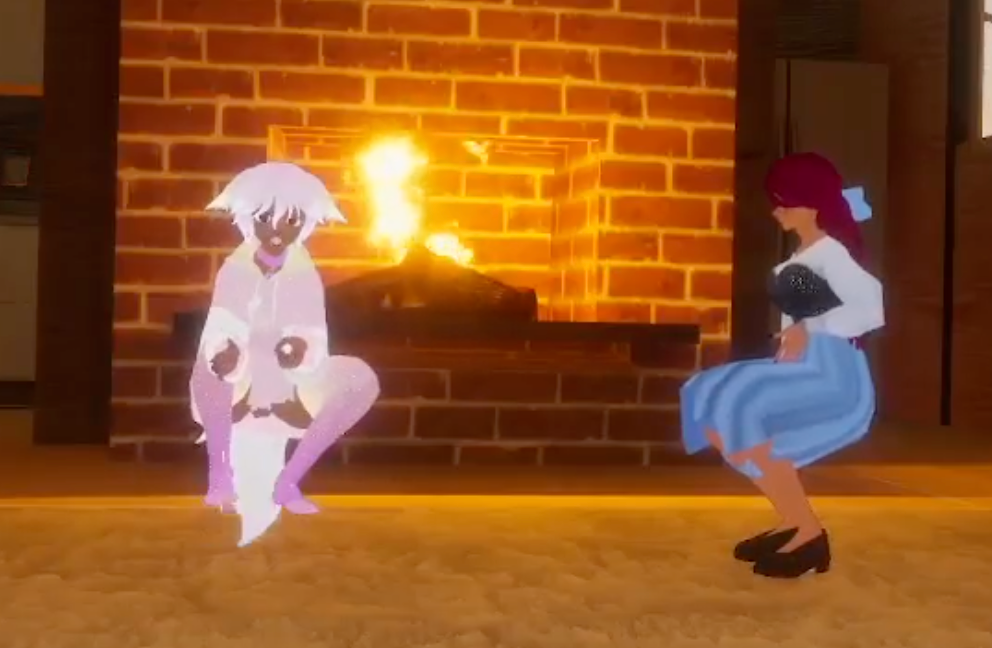}\hspace{4pt}\includegraphics[width=0.48\linewidth,height=4.2cm]{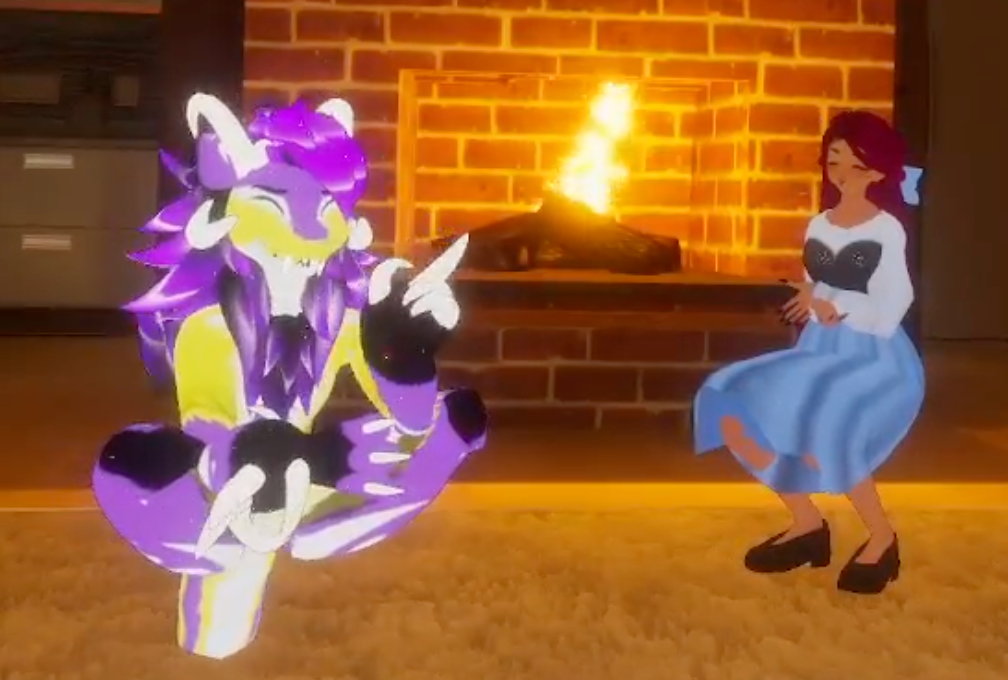}
    \caption{Echo (left) cycling through different avatars during the interview next to Author~1 (right).}
    \Description{Echo cycles through avatars while Author~1 appears alongside in a VRChat environment.}
    \label{fig:case-2}
\end{figure}

Curating my digital persona was intentional, as the platform's virtual affordances gave me agency in shaping how I would initially be perceived through my representation. Ultimately, I decided to choose a human avatar that felt like me: a Black woman. While what ``feels like you'' can vary from person to person, to me, this meant an avatar that reflected core parts of my identity in virtual space. Specifically, I wanted an avatar reflective of my race and gender. Although finding Black, feminine avatars was challenging due to representation gaps on the platform \cite{deveaux_black_2025}, I eventually found one that would become synonymous with how participants would identify me: an avatar of Ariel from the live-action version of \textit{The Little Mermaid} (2023) (See Figure ~\ref{fig:case-2}).

My player arc was similar to what many participants described in their interviews– one that began in public lobbies and progressed toward finding private communities. Unlike private worlds, public worlds are open to anyone who wants to join them. At their best, they offer opportunities for spontaneous adventure and connecting with new players. However, these positive interactions can be overshadowed by griefing, where players intentionally disrupt others’ experiences. Despite being an ethnographer, I was not immune to this. 

I faced embodied racial harassment on multiple occasions. For instance, one day I found myself singing with a group of spontaneous connections in a public world. As one player strummed their guitar through their microphone, we gathered in a circle atop a grassy hill, singing together under a starry, moonlit virtual sky. Mid-song, a user came directly within what felt like my personal (albeit virtual) space and started repeatedly calling me a racial slur. The other singers did not acknowledge the incident. Instead, they appeared to block the offender and carry on. As the one targeted, I was caught between bringing attention to the incident or ignoring it like those around me. Reluctantly, I chose the latter. Browsing through Reddit threads on racism in VRChat further demonstrated an ethos of blocking and moving on. Some of the most liked comments made this recommendation, while others flippantly remarked: ``You ever played an online videogame before?'' Although blocking made the user disappear, the unpleasant feeling it caused did not. I felt disconcerted by the stigmatization of Black digital (and physical) bodies on the platform and came to terms with the fact that my experience as a virtual ethnographer would ultimately be shaped by this reality.

I eventually learned that many adults on VRChat spent most of their time in private lobbies for privacy, safety, and social reasons. To engage in this core aspect of the VRChat experience and mitigate the risk of experiencing harassment, I sought off-platform, online groups dedicated to cultivating communities in VRChat. Mindful of the importance of transparency, I disclosed my researcher identity when I reached out to the moderators of these communities for permission to recruit participants. I also remained cognizant of this in VR, where I stated my role as a researcher in my bio and adaptively navigated ways to ensure players in my vicinity were aware of when and from whom I was collecting data. The nature of the platform sometimes allowed friends of participants to join a room mid-session, adding another layer to maintaining informed consent. Nonetheless, despite the salience of my research identity, I found that my experiences as a VRChat player allowed me to build connections with other players, both fleeting and persistent, that expanded my friend list beyond just research participants. 

Navigating VRChat as someone with intersectional, marginalized identities facilitated some of these connections, allowing me to relate to players with similar experiences. As my research became more focused on racialized experiences, I found an online community dedicated to players of VRChat who came from marginalized backgrounds. This is how I met Echo, another Black VRChat player. Echo selected an indoor location for their interview world, equipped with a fireplace that we sat next to as they shared their VRChat stories (See Figure ~\ref{fig:case-2}). 

During one portion of the interview, Echo showed me their avatars. I watched as Echo switched their appearance from avatar to avatar while expressing how they reflected different aspects of themself. Over time, Echo consistently preferred avatars that were representations of anthropomorphic animals. One was an expressive canine model that smiled when they smiled in real life. What stood out to me about this avatar is that it was one that they took the time to modify. They used technical software to change its color and give it locs, a hairstyle rooted in Black culture. After they shared a desire to see more avatars with characteristics representative of people of color, I asked them what aspects of their identity they wanted to be reflected in their virtual self. To illustrate their thoughts, they switched to an avatar with straight hair. They explained how most avatars had this hair type and how they wished there was more representation of locs, braids, and cornrows. Their challenges with representation on the platform reminded me of my own experience struggling to find Black human avatars. As I sat nodding along to their words, I hoped that my embodied representation and listening ear were a signal of affirmation. It was a sentiment that I aimed to capture further when writing my research findings \cite{deveaux_black_2025}.

The parallelization of play and research, alongside the implications of social VR as an immersive medium, that occurred in this case study reflected emergent relationality. Researching VRChat required engaging with it as a player, an experience that, like that of my participants, was shaped by instances of marginalization tied to my physical identity and its embodied, virtualized expression. Although these challenges were disruptive, they facilitated connections with players who shared similar experiences and inspired new lines of inquiry. In addition, these dynamics shaped how I navigated the platform’s logistics and communities therein.

\subsection*{Of liminal publics on WhatsApp (Author of second story)}
I study caste and gender in computing, and one of my sources of insights is women-in-technology groups. Between 2018-2022, I conducted a total of 25 months of fieldwork with women-in-technology (WIT) groups in the Information Technology (IT) industry in India primarily through Employee Resource Groups in corporations and global conferences. While my first contact with some of the women who were actively organizing and participating in these communities was in-person, in conferences and other communities of support, the onset of Covid-19 fundamentally changed the landscape of life and work – including my approach to doing fieldwork. Between 2020 and 2022, a lot of my contact with my interlocutors was sporadic and primarily mediated through WhatsApp (although platforms like Zoom and Microsoft Teams were also used). Here, I did both participant observation within community events and gatherings online as well as ethnographic interviews with women I met there.

I was primarily interested in knowing how women who are active in WIT groups make sense of their identity as \textit{women} in the industry. I was also interested in knowing what they thought about caste in the computing industry. I observed that caste was usually codified in the group chat. For example, during a prominent Hindu festival, picture and text posts suggesting Hindu goddesses as a source of inspiration for unlocking the potential of modern technologists were circulated. This signaled a form of proximity to theological and scriptural knowledge of Hindu texts that is usually more accessible to the upper castes. The few explicit mentions of caste were done nonchalantly: for example, a person who wrote a book about her experiences as a Kashmiri Brahmin, advertised her book in the group chat framing her Brahmin-ness as a matter of fact. Thus, invocation of caste in both codified and explicit ways in the group normalized it as a sort of a non-issue. In the past seven years that I have been in these groups, caste has never been problematized as a marker/relation of inequality.

In 2018, I was introduced to the India chapter of SheLink, a WIT support group that originated in San Francisco. This chapter had a burgeoning WhatsApp community that was used to announce events, share job opportunities, and seek advice from each other. Richa and I had long been part of this WhatsApp group but were brought together in a smaller circle of members who wanted to organize and offer support through the pandemic: a sub-group called SheLink Wellness group. I was introduced to Richa in a group Zoom call with the other members of this sub-group. 

Richa and I had a couple of interactions in 2020 and 2021, either in events or in planning some group activities aiming to lift the spirits of the women in the group as we navigated a global pandemic. In February 2022, I reached out to Richa just to check in and see how she was doing, and whether she would be open to talking to me about my research on diversity and inclusion in the Indian technology industry. She was going to be one of the first people I interviewed for this research. 
 
Naturally, I had sensed a sort of unease around the explicit mention of caste when I previously brought the concern to Pratibha, another interlocutor I met early on in my research who was a prominent member of the SheLink group, whom I knew much better than Richa. So, to play it safe, I decided to position the question between questions that focused on the category of WIT. Learning to understand the emic and etic categories of the community I was studying, like any ethnographer, I had to experiment with the language, context, and framework to build trusting relationships in this community. The absence of any discussion of caste as an issue of diversity and inclusion in these WIT groups made me feel that broaching the question of caste itself would drive people away from me. I knew it had to be done delicately.

The conversation began with a discussion of her background in terms of family, community, and later in education, specifically in technology. After a deep dive into her experience in the industry and her motivation to work with WIT groups, I asked her about the absence of conversation around the question of caste and religion. I built on this set of structured topics contextually to ask follow-up questions. I asked her if she knew about the Cisco caste-discrimination case\footnote{In a 2019 lawsuit, the State of California accused Cisco Systems of tolerating and being complicit in the ongoing discrimination of a Dalit (marginalized caste) employee of Indian origin by Brahmin (dominant caste) managers \cite{california_court_of_appeal_sixth_appellate_district_department_2022}. Since this lawsuit, several reports of caste-discrimination in the US have emerged \cite{dutt_specter_2020,tiku_googles_2022,rai_how_2021} refuting a long-standing counter-narrative of the technology industry as being `casteless' or meritocratic \cite{fernandez_new_2017}. Several companies like Apple and IBM have since then included caste in their anti-discrimination policy and several civic communities are pushing to make caste a protected category (In 2022, Seattle became the first US city to outlaw caste discrimination).} that had made waves in the US and global technology industry by bringing the issue of caste in such workplaces to the fore. I asked whether she had heard any discussion about the same within her community at work or otherwise. I also adopted a tactic of asking about the parallel of race with caste to see if that might spark some discussions on inclusion and diversity efforts in India's technology industry. Finally, we returned to discussions about who is included in the category of women-in-technology and the reasons for the lack of women in leadership roles. Toward the end, I asked one follow-up question regarding a point she made in our discussion of caste before thanking her for her time and ending the call. In our 90-minute conversation, we had spent 20 minutes talking about her perspectives on caste in the technology industry. 

Fifteen minutes after we ended our call, I received a direct WhatsApp message from her. The message withdrew consent to use her data because she found the conversation to be less about women in technology and more about caste (paraphrased). I responded courteously. I apologized for how she felt and justified my interest in caste as something that was emerging as a potentially interesting dimension of diversity and inclusion efforts in computing. I pointed out its relationship to the idea of “women in technology,” since that was what she wanted to talk about. Unfortunately, I never received a response. 

Two analytical insights from this experience illustrate the dynamic of emergent relationality: i) WhatsApp group chat as a site of legitimizing a collective disavowal of caste, and ii) WhatsApp direct messages as a way of reading critically against such disavowal. 

As outlined above, upper caste members of the community invoke caste explicitly and implicitly. This is done by invoking caste nevertheless as something other than caste \cite{pandian_one_2002} by successfully disavowing it while simultaneously letting it shape the group dynamics. Other members of the SheLink group collectively approve and normalize such invocations to stabilize an upper-caste discourse and practice of diversity and inclusion that disavows caste. This is not to say that all members in the group are from upper caste communities nor does it imply that individually the members of this community do not care about caste, perhaps some do. However, given the public nature of this group, no one (as yet) wanted to disrupt these norms. I too have never raised the ``issue'' of caste in this larger group to avoid alienation from the community. The platform's design contributes to the legitimization of a disavowal of caste and shapes the "publics" the group cares about.

To juxtapose this with my direct message on WhatsApp with Richa, caste was made an \textit{explicit} problem, an issue, because her consent was withdrawn precisely because of explicit questions about caste. Caste could no longer be liminal, making disavowal impossible. Negotiating what is an issue worth raising or acceptable, in this case “women in technology” without a discussion of caste, became a one-on-one, contextual and relational exercise where the actor foreclosed any further elaboration on the topic of caste. My attempts to repair the rupture with Richa through direct messages were futile. My rapport in the larger group setting as someone who cares about “women in technology” as a common issue of concern relied on participating in the culture of normalizing caste conversations and not necessarily making it a problem/issue: I was no longer an “insider” when I explicitly made caste an issue. 

In the normative worlds of computing, where caste is a non-issue, Richa's reaction affirmed that giving space to caste would take away the legitimacy of what had come to be stabilized as matters of ``women in technology.'' Caste could not matter to the identity of women in technology. Richa, a Brahmin woman, revoked her consent because I, a Dalit ethnographer,\footnote{For more on the term Dalit and the position of the untouchable in the caste system see Ambedkar's \cite{ambedkar_untouchables_2008} \textit{Who were the untouchables?} and Rao’s \cite{rao_caste_2009} \textit{The Caste Question}.} elicited a response about caste.\footnote{Although Richa was not aware I am Dalit, it still contributes to my epistemology and analysis of this episode.} In following the IRB protocol, one that is hardly equipped to handle relationships of power in context, I cannot claim the story is anything more than a mere reflection or a fiction of my memory in the field. 

How might a feminist ethnography contextualize the question of refusal where assumptions of mutual trust and solidarity from shared identities (i.e., the collective imaginary of Indian women in computing) are ruptured through disavowed categories like caste? Viswesvaran offers a method that I adopt here: I make the refusal of Richa to be a subject of research \textit{itself} \cite{visweswaran_fictions_1994}. Thus, in line with how Visweswaran evokes fictions of feminist ethnography, I refuse to treat feminist research as unblemished and inherently ethical, then I too like Visweswaran, “ask my subject(s) to pardon me…” \cite [p. 72]{visweswaran_fictions_1994}. 
\section{Discussion}

The observations and reflections in the two case studies show how ethnography is central to the study of evolving social, cultural, and political encounters and issues mediated by hybrid media environments. In the first story, the author navigates dual roles as a Black researcher and social VR player, who in spite of racial abuses in social VR spaces, cultivates kindred connections with other Black players. The second story examines the tensions in the supposedly casteless worlds of computing, highlighting the ways in which the cultural and political stakes of caste are challenged or complicated within women-in-technology groups. Across the hybrid media environments where issues of concern are articulated, forming or dissolving publics, the affective and tactical forms of work performed by ethnographers are captured in the analytic we have termed emergent relationality. This analytic emphasizes situated and interpretive ways of \textit{seeing} and \textit{feeling} that only emerge through ethnographic field practice in hybrid media environments. In the following paragraphs, we discuss two distinct features of emergent relationality: platform/public comprehension and affective/ethical practice. We also offer broader implications for ethnography and design research in CSCW and allied fields.

\subsection{How emergent relationality works}

Emergent relationality is not a method but an analytic composed of registers of observing and attending to the complex and multi-layered aspects of fieldwork, where entangled relations and breakdowns are an ever-present part of practice. In our work, emergent relationality builds on the duality of roles that ethnographers occupy, as observers and participants, as outsiders and insiders, traversing offline-online worlds (what we have identified as hybrid media environments throughout this paper). By emphasizing the tactical and affective work that ethnographers perform, emergent relationality highlights the research skills and improvisations leveraged in examining questions and issues of concern at the interfaces of platforms and publics. Emergent relationality is attuned to worlds that are increasingly mediated through hybrid media environments, and how they influence and extend expression, belonging, and deliberation. 

The main goal of this analytic is to support the examination of issues that emerge in everyday interactions with(in) hybrid media environments. As shown in our research, where there is an issue, there is a \textit{way} -- a way through which a public(s) emerges. Issues are a necessary condition for the formation of a public(s), but do not solely determine its constitution and resolution. The making of publics critically depends on the \textit{way} relations are forged across hybrid media environments, including how norms and worldviews of class, gender, and race are articulated (or silenced). Publics are additionally shaped and influenced through aspects of individuals or groups that are made visible or performed publicly. This orientation to issues (hence publics) is constituted in infrastructural participation \cite{baringhorst_how_2019,marres_infrastructural_2025,korn_infrastructuring_2019}, where the role of ethnographers as active participants cannot be understated. What might ethnography offer to the study of publics in the digital age? In the following subsections, we show the two features of emergent relationality: platform/publics comprehension and affective/ethical praxis.

\subsubsection{Emergent relationality as platform and publics comprehension}
Apprehending how platforms and publics are constituted and entangled is an important aspect of field practice. Across the two case studies in this paper, platforms – i.e., VRChat and WhatsApp –  were central not only to the ways we connected and wove networks across field sites but also constitutive of the themes and lines of inquiry we explored, developed, and subsequently reflected upon. In the first story, the modalities of voice and spatial awareness in VRChat produced engaging and immersive interactions, which helped the author connect with her interlocutors (although similarly exposed her to racial abuse). In the second story, an understanding of what remained unspoken or a non-issue in the (larger) public WhatsApp group and eventually a matter of breakdown over a (private) direct message revealed the practical and discursive limits of caste in moderating access to women-in-technology groups.

By paying attention to the mutual constitution of publics and platforms, our field sites were constructed as always entangled and always in flux, demanding ongoing work in situating ourselves relative to participants, platforms, and issues of concern. Building on existing work that argues for publics as overlapping, multiple, and emergent \cite{starr_relational_2021,dewey_public_2016}, our research shows that publics are both the {\em name} of the issues that emerge in situated practice and a {\em site} through which these issues are articulated and negotiated. Relations of race and caste are not only made explicit (by the researchers) but are also seen as a source of insight into what counts as an issue/concern (and therefore publics). 

However, the comprehension of platforms and publics advocated here goes beyond the necessary work of auditing and exposing how platforms function \cite{eslami_i_2015,karizat_algorithmic_2021} or how publics are infrastructured in online \cite{boyd_social_2011,gillespie_relevance_2014,bonilla_ferguson_2015,tufekci_engineering_2014} and hybrid media contexts \cite{moller_hartley_researching_2023,willems_politics_2019}. It goes beyond the complexities and challenges of content moderation on platforms, including social messaging \cite{shahid_one_2025,agarwal_conversational_2024,varanasi_accost_2022} and social VR \cite{sabri_challenges_2023}. Rather, we wish to underscore that what the ethnographer understands about the platform (affordances, discourse, modalities, policies, etc.) and the ethnographer’s relative position to the platform (and their subjects of study) is critical to this comprehension. Similarly, the work of comprehending the relationship between platforms enables ethnographers to appreciate their mutual shaping in a more practical albeit performative sense. For instance, this work can reveal the discursive and material ways that actors (observers and participants) aim to maintain the stability of social and interactional orders, and ultimately what platforms amplify or diminish. By focusing on such performances, ethnographers are better equipped to handle concerns about transparency and research reliability. For the first story, this meant that the ethnographer/author was able to find practices of solidarity and care (on VRChat), not by \textit{understanding} the inner workings of the platform but by observing the norms and values VRChat users cultivated for safety. For the second story, the ethnographer had to broach a discussion of caste while also navigating the risk of losing field access and modalities of “fitting in” as a participant (on WhatsApp). 

The comprehension of platforms and publics varied across our case studies. While our work builds on relational publics -- “open-ended networks of actors (i.e., without a closed or fixed membership) linked together through flows of communication, shared stories, and civic or other collective concerns” \cite[p. ~69] {starr_relational_2021} -- we approached our sites with less normative baggage and with a critical stance toward the emergent and {\em personal}. Our case studies read \textit{into} and \textit{against} the grain of our relationships with field participants as well as the platforms through which different forms of publics emerged. In the first story, the publics were embodied and reparative, oriented toward belonging and solidarity; in the second story, the publics were liminal, a non-issue in \textit{public} but silenced and disavowed in private.

\subsubsection{Emergent relationality as affective and ethical praxis}
Ethnographic practice is an ongoing affective and ethical process; it would be a fallacy to frame it as an objective process and the experience itself evokes emotions that continue to be questioned and debated by the ethnographer even after they have `left' the field.\footnote{The idea that one ever `leaves' the field is perhaps also a fallacy.} For example, after experiencing racial harassment on VRChat, the author sought advice from Reddit (and elsewhere) on how to deal with her unenviable predicament. Although she could have remained anonymous on social VR, honing Black avatars and Black forms of expression concretized her search towards the Black experiences of belonging within and beyond social VR. Paying attention to the experience of racial hostility was integral to seeking a network of shared affinities, joys, and aspirations. An alternative kind of publics emerged, not by dint of imagination but rather through the emotional labor of working through adversity and finding a kindred community. 

Questions around membership and access shaped our work across the two stories, drawing on reflexive practices common in CSCW and HCI while also challenging them. In the second story, the ethnographer spent several years establishing connections and building rapport with women in technology (within the bounds of her university's IRB requirements). Yet honoring the ethical commitment to represent caste perspectives sometimes required risking insider status, and even then, tensions and breakdowns were common. The affective work of restoring order took the form of ongoing negotiations of access. The affective work performed to restore order was through ongoing negotiations of access. That is why the rupture of refusal in Richa's case itself became the subject of research, prompting reflection on navigating women in technology spaces as an ``outsider-within'' \cite{collins_reflections_1999}. Across our case studies, we also inverted the trope of studying the ‘native’ by studying up \cite{nader_up_1972,gusterson_studying_1997} or by studying {\em across} communities embedded in race and caste, within varied axes of access and visibility, while confronting sociotechnical harms and disavowals.

\subsection{Implications for methods}
Our work is by no means new. Several issues we discussed and reflected upon have been of concern since the foundation of CSCW, and have undergone change and renewed thinking in the recent past. However, we offer a common grammar for the entanglement of relations between ethnographers, platforms, and publics, and an analytic that can help articulate how these relations are practically and ethically navigated. 

But what does this mean for (digital) ethnography? In this paper, while we have blurred the distinctions between digital ethnography and classic ethnography, we do not suggest that these modes of ethnography are the \textit{same}. In fact, ethnographies vary depending on disciplinary commitments and intellectual traditions \cite{bell_problem_2019,rode_commentary_2026}. However, infrastructural participation has radically changed how we experience space and the issues actors articulate and coalesce around. Despite considerable changes in `sites' of focus in CSCW – from traditional workplaces to groupware and virtual environments – \citet{dourish_re-space-ing_2006} argues that these apparent transformations have not altered CSCW’s sensitivity to space and context. 

Indeed, space and context (and action) matter in understanding key aspects of ethnography in the digital age, insofar as they facilitate or foreclose participant observation. Although technology in many ways forces us to re-encounter and re-imagine social and cultural spaces, these spaces are experienced within the contexts actors find meaningful \cite{dourish_re-space-ing_2006}. Action is situated. Yet participant observation, as we know it, has radically changed with these shifts in technology across time. The temporal rhythms of ethnography as an immersive and real-time process (and product) have changed because ethnographers cannot be “always on.” As such, participant observation becomes more polymorphous and distributed \cite{gusterson_studying_1997,marcus_ethnography_1995}, with ethnographers {\em collaborating} closely with archives and databases, supplemented by various research interventions – e.g., collaborative (auto)ethnography \cite{chang_collaborative_2013,lassiter_chicago_2005}, participation action research\cite{hayes_relationship_2011}, etc. – that cut across the online and real world offline contexts. Together, ethnography echoes the hybridity that characterizes the messy worlds of data and algorithms \cite{seaver_bastard_2015}, in which “thick description” is no longer only about the “webs of significance” \cite{geertz_thick_1973} that actors (including participants and ethnographers) spin, but also about the \textit{webs and networks} through which these actors themselves are spun by complex assemblages of data, algorithms, and user activity, as well as platform logics.

Our work also has implications for design research methods, which we situate in relation to ethnography with particular focus on publics (although our takeaways can be applied to \textit{any} unit of analysis besides publics). While ethnographic research and design of publics has been of major interest to CSCW, HCI, and design studies, each of these disciplines tend to treat publics differently. For ethnography, publics are a unit of analysis that can be observed and interpreted in their transformations \cite{coleman_hacker_2011,metcalf_experimental_2025,ludwig_publics_2016}; whereas for HCI and design, publics are a latent \textit{quality} with some form of political potential that can be actualized through design intervention, fostering social change around issues of concern \cite{disalvo_design_2009,disalvo_making_2014,dantec_infrastructuring_2013}. However, the participatory processes that facilitate action across ethnographic and design practice open spaces for new forms of inquiry into the infrastructures that support social and political action in the world. 

Emergent relationality resonates with design research approaches, including speculative design and Research through Design \cite{zimmerman_analysis_2010}. In these approaches, solutions are not the ultimate product; instead, attention is directed toward generating new questions and knowledge that reframe how problems are understood. Since design research (and other adjacent fields) take seriously positionality (hence subjectivity) as well as scale (complexities that are difficult to unravel) \cite{beckwith_scale_2020,joshi_who_2024,khovanskaya_reworking_2017}, emergent relationality cultivates productive friction for example between designers and ethnographers, allowing each role to leverage its strengths without flattening their differences \cite{khovanskaya_reworking_2017}. This is particularly important in increasingly hybrid media environments where the boundaries between the real, virtual, and the in-between are increasingly blurred. Researchers must develop the tactics and sensibilities to \textit{see} and \textit{feel} what it means to be on the other end, including exposure to real harms or opportunities for community-building (as we have seen in our case studies). Where there is an issue, there is a \textit{way}.

\section{Conclusion}

The question posed in the paper's title -- \textit{To Tango or to Disentangle?} -- sits at the heart of ethnography and qualitative research of the hybrid online-offline spaces and communities. In this paper, we focus on the unexpected and emergent forms of relations and social practices that entangle ethnographers and their sites and subjects of study. Extending the analysis from the duality of roles occupied by ethnographers (at once observers and participants, outsiders and insiders), the paper introduces emergent relationality as a tripartite analytic for understanding the emergent relationships and practices at the intersection of platforms, publics, and ethnography. We examined how positionality and hybrid media environments constitute and condition access and participation and ultimately how issues of concern are articulated and negotiated. In the two stories developed, focusing on issues of race and caste on VRChat and WhatsApp, respectively, we explored how the situated practices through which ethnographers manage field performances lend themselves to rich understandings of platforms and the publics they cultivate. We read into how experiences of Blackness on VRChat exposed actors to real harm, while opening spaces for belonging. At the same time, we read into the disavowal of caste within the supposed programs of solidarity and support in the worlds of women in computing. 
It is against this work that emergent relationality provides the nuanced lens researchers and reviewers (within and beyond CSCW) need to pursue and evaluate ethically and rigorously \textit{pragmatic} ethnography of people, platforms, and publics in the digital age.


\begin{acks}
We thank Madiha Z. Choksi for contributions during earlier stages of this project, and Pegah Moradi and Aspen Omapang for feedback on earlier versions of this paper. We also appreciate the instructive and generous guidance from the anonymous reviewers. Any errors or omissions are our own.
\end{acks}

\bibliographystyle{ACM-Reference-Format}
\bibliography{references}

\received{May 2025}
\received[revised]{November 2025}
\received[accepted]{December 2025}

\end{document}